\documentclass[12pt]{article}
\def\be{\begin{equation}}
\def\ee{\end{equation}}
\def\ba{\begin{eqnarray}}
\def\ea{\end{eqnarray}}
\def\nn{\nonumber}
\def\lb{\label}
\def\bb{\bibitem}
\def\dfrac{\displaystyle\frac}
\def\C{{\cal C}}
\begin{document}



\begin{titlepage}
\date{}
\title{
\begin{flushright}\begin{small} LAPTH-065/23 \end{small} \end{flushright} \vspace{1cm}
Cotton gravity is not predictive}



\author{G\'erard Cl\'ement$^a$\thanks{Email: gclement@lapth.cnrs.fr} and
Khireddine Nouicer$^{b}$\thanks{Email: khnouicer@univ-jijel.dz} \\ \\
$^a$ {\small LAPTh, Universit\'e Savoie Mont Blanc, CNRS, F-74940  Annecy, France} \\
$^b$ {\small LPTh, Department of Physics, University of Jijel,} \\
{\small BP 98, Ouled Aissa, Jijel 18000, Algeria}}

\maketitle

\begin{abstract}
It is well known that the theory of Cotton gravity proposed by Harada is trivially solved by all isotropic and homogeneous cosmologies. We show that this under-determination is more general. More precisely, the degree of arbitrariness in the solutions increases with the degree of symmetry. The inverse problem of finding the matter source generating a given spacetime geometry is similarly indeterminate in the case of highly symmetric configurations. We give two simple examples. The first is that of static spherically symmetric solutions, which depend on an arbitrary function of the radial coordinate. The second is that of anisotropic cosmologies, which depend on an arbitrary function of time.
\end{abstract}

\end{titlepage}
\setcounter{page}{2}

\section{Introduction}
Despite its mathematical elegance and its ability to describe phenomena on the scale of the solar system and beyond, the theory of general relativity has encountered observational challenges at the cosmic level. These have been solved by the introduction of new paradigms such as dark matter and dark energy \cite{will_2014}. Currently, neither general relativity itself nor the standard model of particle physics have succeeded in providing convincing explanations for these new fundamental concepts. These considerations have led to the development of a plethora of extended theories of gravity. In this paper we will focus on the theory recently proposed by J. Harada and dubbed Cotton gravity\footnote{Not to be confused with the work by F.W. Cotton, ``A generalisation of the Einstein-Maxwell equations'', Eur. Phys. J. Plus 136 (2021) 162.} \cite{Harada_2021}. In this theory the gravitational dynamics are described, not by the Einstein tensor, but by the rank-three Cotton tensor \cite{cotton_1899}. By construction, any solution of the general relativity with or without a cosmological constant is also a solution of Cotton gravity, the cosmological constant arising as an integration constant upon solving the third-order field equations. The first Schwarzschild-like non-trivial vacuum solution of Cotton gravity was derived in \cite{Harada_2021}. In a subsequent paper, J. Harada showed that the galactic rotation curves can be explained without the need for dark matter contribution \cite{Harada_2022}. Harada's spherically symmetric solution was generalized in \cite{gogberashvili2023general}. More recently, using a parametrization of Cotton gravity based on Codazzi tensors, first introduced in \cite{Mantica_2023}, a set of new solutions of Cotton gravity were discussed in \cite{Mantica_20232,sussman2023cotton,sussman2023exact}.

The theory of Cotton gravity was first challenged by Bargue\~no \cite{Bargueno_2021}, who argued that this theory had less information than general relavity. The arguments of \cite{Bargueno_2021} were not conclusive and were refuted by Harada in \cite{Harada_2021R}. On the other hand, it is well known that the Cotton tensor for conformally flat Riemannian manifolds vanishes identically \cite{garcia_2004}, so that all Friedman-Lema\^{\i}tre-Robertson-Walker (FLRW) spacetimes solve trivially the equations of Cotton gravity. In this letter, we show that the problem with this theory goes beyond the case of isotropic and homogeneous spacetimes. Specifically, we shall see that the degree of arbitrariness in the solutions of Cotton gravity increases with the degree of symmetry assumed. Such an indetermination has previously been observed in other modified gravity theories, such as generalized teleparallel gravity \cite{isuzi2014}.

After briefly presenting the theory of Cotton gravity, we treat in the following, first the case of static spherically symmetric solutions, which we show to depend on an arbitrary function of the radial coordinate, then the case of anisotropic cosmologies, which depend on an arbitrary function of time. We also discuss the Codazzi tensor approach to Cotton gravity, which we  show to suffer from the same indetermination.

\section{Cotton gravity}
The field equations of Cotton gravity \cite{Harada_2021} can be written as
\be\lb{eqcott}
C_{\mu\nu\rho} = 8\pi G M_{\mu\nu\rho},
\ee
where
\be\lb{defcott}
C_{\mu\nu\rho} \equiv D_\mu R_{\nu\rho} - D_\nu R_{\mu\rho} - \dfrac16\left(g_{\nu\rho}D_\mu R - g_{\mu\rho}D_\nu R\right)
\ee
(with $R_{\mu\nu}$ the Ricci tensor for the metric $g_{\mu\nu}$, $R$ its trace, and $D_\mu$ the covariant derivative) is the Cotton tensor, and
\be\lb{gam}
M_{\mu\nu\rho} \equiv D_\mu T_{\nu\rho} - D_\nu T_{\mu\rho} - \dfrac13\left(g_{\nu\rho}D_\mu T - g_{\mu\rho}D_\nu T\right)
\ee
(with $T_{\mu\nu}$ the matter energy-momentum tensor) is the generalized angular momentum tensor. The Cotton tensor is subject to the algebraic identities
\ba
&& C_{\mu\nu\rho} + C_{\nu\mu\rho} \equiv 0 , \lb{antisym} \\
&& C_{\mu\nu\rho} + C_{\nu\rho\mu} + C_{\rho\mu\nu} \equiv 0 , \lb{circ} \\
&& g^{\nu\rho} C_{\mu\nu\rho} \equiv 0 . \lb{bianchi}
\ea
The first identity (antisymmetry in the first two indices) results directly from the definition (\ref{defcott}), the second results from the symmetry of the Ricci and metric tensors, and the third embodies the Bianchi identity for the Ricci tensor. The matter tensor $M_{\mu\nu\rho}$ is subject to similar identities, the last one
\be\lb{conserv}
g^{\nu\rho} M_{\mu\nu\rho} = 0 .
\ee
resulting from the conservation law of the energy-momentum tensor.

In the case of a generic four-dimensional spacetime, the Cotton tensor has 16 algebraically independent components \cite{garcia_2004}. So it would seem that the equations of Cotton gravity are over-determined, the metric tensor having only 10 algebraically independent components. However, any solution of the Einstein equations with cosmological constant
\be
G_{\mu\nu} + \Lambda g_{\mu\nu} = 8\pi G T_{\mu\nu}
\ee
solves the equations of Cotton gravity. Accordingly, these do have a solution, up to some integration constants. At this point, let us recall that the 10 Einstein equations are not functionally independent, as the components of the Einstein tensor $G_{\mu\nu} = R_{\mu\nu} - (1/2)R g_{\mu\nu}$ are related by the 4 Bianchi identities $D_\nu{G_\mu}^\nu = 0$, so that the metric tensor components $g_{\mu\nu}$ are determined only up to 4 arbitrary coordinate transformations (gauge conditions) \cite{Weinberg_1972}.

\section{Spherical symmetry}

Let us now assume the existence of two commuting Killing vectors. This is the case of stationary axisymmetric metrics, which may be parameterized in the Weyl form by
\be
ds^2 = g_{ab}(\rho,z)\,dx^a\,dx^b + e^{2\nu(\rho,z)}(d\rho^2+dz^2) \quad ({\rm det}(g_{ab})=-\rho^2)
\ee
($a,b=0,1$ with $x^0=t$, $x^1=\varphi$). Using the block-diagonal character of this metric and the independence of the metric functions on the cyclic coordinates $x^a$, it is easy to see that the components $C_{abc}$ and $C_{iaj}$ ($i,j= 2,3$ with $x^2=\rho$, $x^3=z$), as well as $C_{aij}$ and $C_{ija}$, vanish identically. Furthermore, the two-space $(\rho,z)$ is conformally flat, so that its Cotton tensor
components $C_{ijk}$ also vanish. The only non-vanishing Cotton tensor components are thus the 6 $C_{aib}$ (the other mixed components $C_{iab}$
and $C_{abi}$ are related to these by (\ref{antisym}) and (\ref{circ})) which are related by the 2 Bianchi identities (\ref{bianchi})
\be
g^{ab} C_{aib} = 0,
\ee
leaving $6-2=4$ independent Cotton tensor components. Again, the 4 independent equations of Cotton gravity are more than enough to determine the 3 independent metric tensor components (two components of $g_{ab}$ and $ e^{2\nu}$).

Now let us enhance the symmetry and consider static spherically symmetric metrics (four Killing vectors, two of which commute).
We use the standard parametrization with areal radial coordinate:
\be\lb{SS}
ds^2 = - e^{2\nu(r)}\,dt^2 + e^{2\lambda(r)}\,dr^2 + r^2(dr^2 + \sin^2\theta\,d\varphi^2).
\ee
The non-vanishing components of the Ricci tensor are the diagonal $R_{tt}$, $R_{rr}$, $R_{\theta\theta}$ and $R_{\varphi\varphi} = \sin^2\theta\, R_{\theta\theta}$. Using the diagonal character of the metric and the independence of the metric functions on the cyclic coordinates $t$, $\theta$ and $\varphi$, one finds that the only independent non-vanishing Cotton tensor components are $C_{trt}$, $C_{\theta r\theta}$ and $C_{\varphi r\varphi} = \sin^2\theta\,C_{\theta r\theta}$, the other non-vanishing components $C_{rtt}$, $C_{ttr}$, etc. being related to these by (\ref{antisym}) and (\ref{circ}). So apparently we have two independent Cotton tensor components. But again, these are related by the Bianchi identity (\ref{bianchi})
\be\lb{BS}
g^{tt}C_{trt} + g^{\theta\theta}C_{\theta r\theta} + g^{\varphi\varphi}C_{\varphi r\varphi} = g^{tt}C_{trt} + 2\,g^{\theta\theta}C_{\theta r\theta} = 0.
\ee

So in this case the Cotton tensor has only one non-vanishing independent component. As the matter tensor is subject to the same algebraic identities (\ref{antisym}), (\ref{circ}) and (\ref{bianchi}), it follows that the equations (\ref{eqcott}) of Cotton gravity reduce in the static spherically symmetric case to only one equation for the two metric functions $e^{2\nu(r)}$ and $e^{2\lambda(r)}$, which reads in the vacuum case ($T_{\mu\nu}=0$, implying $M_{\mu\nu\rho} = 0$):
\ba\lb{CS}
C_{trt} &\equiv& \dfrac23\,e^{2(\nu-\lambda)}\left\{\nu''' + \left(2\nu'-3\lambda'+ \dfrac2r\right)\nu'' + \left(-\nu'+\dfrac1r\right)\lambda'' - 2\nu'\lambda'(\nu'-\lambda') \right. \nn\\
&& \left.  + \dfrac1r(\nu'-\lambda')(3\nu'+2\lambda') - \dfrac2{r^2}\nu' + \dfrac1{r^3}(1-e^{2\lambda})\right\}  = 0,
\ea
where $'{} = d/dr$. This can be solved up to one arbitrary function of the radial coordinate and several integration constants. Harada in \cite{Harada_2021} used the Schwarzschild ansatz with $e^{2\lambda} = e^{-2\nu}$, leading to a vacuum solution depending on three integration constants $M$, $\Lambda$ and a new constant $\gamma$. The authors of \cite{gogberashvili2023general} raised this restriction, but chose $e^{2\nu(r)} = e^{2\nu(r)}_{Harada}$, and so obtained for $e^{2\lambda(r)}$ a solution depending on the Harada constants as well as on two new constants $C_1$ and $C_2$, which they claimed was the general static spherically symmetric vacuum solution of Cotton gravity. However it is clear that sourceless Cotton gravity admits a non-denumerable infinity of static spherically symmetric solutions depending on an arbitrary function of $r$.

Consider now the case of a static spherically symmetric matter source. This is characterized by a diagonal energy-momentum tensor $T_{\mu\nu}$, which may be thought of as a static anisotropic fluid, with components
\be\lb{TS}
T_\mu^\nu = {\rm diag}\left(-\rho, p-2\Pi, p+\Pi, p+\Pi\right).
\ee
The conservation law (\ref{conserv}) constrains this tensor by
\be\lb{COS}
p' - 2\Pi' + (\rho+p-2\Pi)\nu' - \dfrac{6\Pi}r = 0,
\ee
leaving at most two independent components of $T_{\mu\nu}$. Similarly to the Cotton tensor, the generalized angular momentum tensor (\ref{gam}) has only one non-vanishing independent component, which may be chosen to be
\be\lb{MS}
M_{trt} = - 2e^{2\nu}\left[\frac13\rho' + \Pi' + \dfrac{3\Pi}r\right].
\ee
It is clear that, for any given functions $\rho(r)$ and $\Pi(r)$, the differential equation for two unknown functions $\nu(r)$ and $\lambda(r)$
\be\lb{CMS}
C_{trt} = 8\pi G M_{trt}
\ee
(with $C_{trt}$ given by the left-hand side (\ref{CS}) and $M_{trt}$ given by (\ref{MS})) admits as in the vacuum case an infinity of solutions depending on an arbitrary function of $r$. Moreover, the inverse problem of finding the matter source generating a given static spherically symmetric spacetime geometry is also indeterminate. For any given metric functions $\nu(r)$ and $\lambda(r)$, equations (\ref{CMS}) and (\ref{COS}) determine the energy-momentum tensor (\ref{TS}) only up to one arbitrary function of $r$.

\section{Anisotropic cosmologies}
Another example is that of homogeneous anisotropic cosmologies, with three commuting Killing vectors. These can be parameterized by
\be\lb{ani}
ds^2 = - dt^2 + a_1^2(t)\,dx^2 + a_2^2(t)\,dy^2 + a_3^2(t)\,dz^2 .
\ee
It is clear from the diagonal form of this metric depending only on the time coordinate and from the algebraic identities (\ref{antisym}) and
(\ref{circ}) that the Cotton tensor (\ref{defcott}) has only three independent components $C_{xtx}$, $C_{yty}$, $C_{ztz}$. However these are related by the Bianchi identities (\ref{bianchi}),
\be
a_1^{-2}C_{xtx} + a_2^{-2}C_{yty} + a_3^{-2}C_{ztz} \equiv 0.
\ee
The components of the matter tensor $M_{\mu\nu\rho}$ being related by the corresponding equation (\ref{conserv}), it follows that the general equations of Cotton gravity reduce in this case to only two independent equations for the three unknown functions $a_1(t)$, $a_2(t)$, $a_3(t)$.
Putting $a_i(t) = e^{\nu_i(t)}$ ($i=1,2,3$), these equations read, in the vacuum case,
\be\lb{anicosm}
\dot{\ddot{\nu_i}} + \dot{\nu_i}\ddot{\nu_i} + [\dot{\nu}]\ddot{\nu_i} + [\dot{\nu}](\dot{\nu_i})^2 - [\dot{\nu}^2]\dot{\nu_i} =   \dfrac13\left([\dot{\ddot{\nu}}] + [\dot{\nu}\ddot{\nu}] + [\dot{\nu}]\,[\ddot{\nu}]\right)
\ee
where $\dot{} = d/dt$, and $[\nu] = \sum_{i=1}^{3}\nu_i$. Again, this under-determination carries over to the case of matter sources with an energy-momentum tensor consistent with the assumed symmetry (e.g. anisotropic perfect fluid), for which the equations of Cotton gravity lead to a metric tensor (\ref{ani}) depending on an arbitrary function of time. When the symmetry is further enhanced to that of isotropic cosmologies (six Killing vectors),
\be
ds^2 = - dt^2 + a^2(t)\,d{{\vec x}}^2 ,
\ee
there remains an arbitrary function of time, the overall scale factor $a(t)$, confirming the well-known fact that the FLRW spacetime is conformally flat, so that its Cotton tensor vanishes identically.

We emphasize again that the under-determination of the equations of Cotton gravity for the symmetric configurations discussed above applies whatever the distribution of matter source consistent with the specified symmetries. The deep reason behind this absence of predictivity of the theory  is that Cotton gravity does not couple directly to the energy-momentum tensor, but only to the generalized angular momentum tensor (\ref{gam}), which carries less information, the information content carried by the generalized angular momentum tensor decreasing when the symmetry of the matter source increases.

\section{Codazzi parametrization}

The authors of \cite{Mantica_2023} suggested to reformulate Cotton gravity in the following manner. Given a matter energy-momentum tensor $T_{\mu\nu}$ and a spacetime metric $g_{\mu\nu}$ with Ricci tensor $R_{\mu\nu}$, not necessarily solving the Einstein equations with source $T_{\mu\nu}$, define a symmetric 2-tensor $\C_{\mu\nu}$ of trace $\C$ by
\be\lb{defcod2}
C_{\mu\nu} \equiv R_{\mu\nu} - \dfrac16 R\,g_{\mu\nu} - 8\pi G \left[T_{\mu\nu} - \dfrac13 T\,g_{\mu\nu}\right].
\ee
This definition can equivalently be reformulated as the generalized Einstein equations
\be\lb{defcod}
R_{\mu\nu} - \dfrac12 R\,g_{\mu\nu} \equiv 8\pi G \left[T_{\mu\nu} + \C_{\mu\nu} - \C g_{\mu\nu}\right],
\ee
with a source combining the contribution of the matter energy-momentum tensor and that of the tensor $\C_{\mu\nu}$, which might correspond to the dark sector of cosmology \cite{Mantica_20232}.

If the Cotton gravity equations (\ref{eqcott}) are satisfied, the tensor $C_{\mu\nu}$ defined by (\ref{defcod2}) fulfils the condition:
\be\lb{charcod}
E_{\mu\nu\rho} = 0,
\ee
where the 3-tensor $E_{\mu\nu\rho}$ is defined by
\be\lb{defE}
E_{\mu\nu\rho} \equiv D_\mu \C_{\nu\rho} - D_\nu \C_{\mu\rho}.
\ee
The condition (\ref{charcod}) on the tensor defined by (\ref{defE}) is the characteristic property of Codazzi tensors. It has been suggested in \cite{sussman2023cotton} that such a ``Codazzi parametrization'' -- defining the tensor $\C_{\mu\nu}$ by (\ref{defcod2}) and then enforcing the Codazzi condition (\ref{charcod}) -- could be used to generate non-trivial FLRW solutions of Cotton gravity. This approach was also extended in \cite{sussman2023exact} to the case of some inhomogeneous and anisotropic cosmologies, as well as to some spherically symmetric spacetimes.

Inserting the definition (\ref{defcod2}) in the definition (\ref{defE}) leads to the equivalent form
\be
E_{\mu\nu\rho} \equiv C_{\mu\nu\rho} - 8\pi GM_{\mu\nu\rho}.
\ee
Thus the Codazzi condition (\ref{charcod}) is completely equivalent to the fundamental equation (\ref{eqcott}) of Cotton gravity. For the sake of clarity we shall now refer to the original formulation of Cotton gravity by equations (\ref{defcott}) and (\ref{eqcott}) as the Harada formulation, and to that by (\ref{defcod2}) and (\ref{charcod}) as the Codazzi formulation. It should be clear that these are two equivalent formulations of the same theory. Owing to this equivalence, the under-determination of the equations of the Harada formulation of Cotton gravity for highly symmetric configurations carries over to the under-determination of the Codazzi formulation for the same configurations.

Recalling that the components of the Cotton tensor $C_{\mu\nu\rho}$ are related by the algebraic identity (\ref{bianchi})
and that the matter tensor $M_{\mu\nu\rho}$ must obey the conservation law (\ref{conserv}), it follows that the components of the tensor $E_{\mu\nu\rho}$ are related by
\be\lb{constr}
g^{\nu\rho}E_{\mu\nu\rho} = 0,
\ee
independently of the Codazzi condition (\ref{charcod}). This constraint lowers the number of independent Codazzi equations, which for highly symmetric configurations becomes smaller than the number of independent metric functions, so that these cannot be determined completely.

Consider again the example of static spherically symmetric configurations (\ref{SS}). Owing to the spherical symmetry of both the metric and the matter source, the non-vanishing components of the tensor $\C_{\mu\nu}$ defined by (\ref{defcod2}) are diagonal, the angular components being related by $\C_{\varphi\varphi} = \sin^2\theta\, \C_{\theta\theta}$. Clearly the radial component $\C_{rr}$ do not contribute to the Codazzi condition (\ref{charcod}), so that there are only two non-trivial Codazzi equations $E_{trt} = 0$ and $E_{\theta r\theta} = 0$. However, these are not independent, because of the constraint (\ref{constr}), which reads here
\be
g^{tt}E_{trt} + 2\,g^{\theta\theta}E_{\theta r\theta} = 0.
\ee
It follows that there is only one Codazzi equation for two unknown metric functions of $r$, which therefore can be determined only up to one arbitrary function of $r$. We could proceed similarly with the cases of homogeneous anisotropic or isotropic cosmologies. At best, the Codazzi parametrization can be used only to classify the Friedmann equations of extended theories of gravity \cite{sussman2023exact,Mantica_20233} or models of spherical configurations of matter \cite{sussman2023exact}.

\section{Conclusion}

We have found that Cotton gravity not only has more solutions than general relativity, but it also has too many solutions, with the degree of arbitrariness increasing with the degree of symmetry. Specifically, the general spherically symmetric solution of Cotton gravity's equations depends on an arbitrary function of the radial coordinate. Similarly, the general anisotropic Cotton gravity cosmology depends on an arbitrary function of time. This under-determination affects both the direct problem of finding the spacetime geometry generated by a given matter source and the inverse problem of finding the matter source generating a given spacetime geometry, particularly in highly symmetric configurations. This issue persists in the alternative Codazzi parametrization of Cotton gravity. Therefore, regardless of the formulation, Harada or Codazzi, Cotton gravity cannot qualify as a physical theory. However, the alternative theory of gravity recently proposed by Harada \cite{Harada_2023}, which also involves differential field equations of the third order, should be free from this shortcoming.

After this paper was submitted for publication, a preprint appeared \cite{justin2024} softening our conclusions. In brief, the conclusion of \cite{justin2024} is that highly symmetric configurations are exceptional, and that the Codazzi tensor $\C_{\mu\nu}$ trivializes to $\lambda g_{\mu\nu}$ on generic backgrounds.

\section*{Acknowledgments}

One of us (K.N.) thanks the LAPTh Annecy, France, for warm hospitality and the University of Jijel, Algeria, for financial support. We are grateful to J. Harada and A. Barnes for valuable comments.



\end{document}